\documentclass[prl,twocolumn,floatfix,superscriptaddress, longbibliography]{revtex4-1}
\usepackage[pdftex,plainpages=false,colorlinks=true,linkcolor=blue, citecolor=blue, urlcolor=blue]{hyperref}
\usepackage{amsfonts}
\usepackage{amsmath}
\usepackage{amssymb}
\usepackage{graphicx}
\usepackage{natbib}
\usepackage{color}
\usepackage{placeins}
\usepackage{physics}

\begin{document}

\title{Specific heat maximum as a signature of Mott physics in the two-dimensional Hubbard model}
\author{G. Sordi}
\email[corresponding author: ]{giovanni.sordi@rhul.ac.uk}
\affiliation{Department of Physics, Royal Holloway, University of London, Egham, Surrey, UK, TW20 0EX}
\author{C. Walsh}
\affiliation{Department of Physics, Royal Holloway, University of London, Egham, Surrey, UK, TW20 0EX}
\author{P. S\'emon}
\affiliation{Computational Science Initiative, Brookhaven National Laboratory, Upton, NY 11973-5000, USA}
\author{A.-M. S. Tremblay}
\affiliation{D\'epartement de physique \& Institut quantique, Universit\'e de Sherbrooke, Sherbrooke, Qu\'ebec, Canada J1K 2R1}
\affiliation{Canadian Institute for Advanced Research, Toronto, Ontario, Canada, M5G 1Z8}

\date{\today}

\begin{abstract}
Recent experiments on cuprates show that as a function of doping, the normal-state specific heat sharply peaks at the doping $\delta^*$, where the pseudogap ends at low temperature. This finding is taken as the thermodynamic signature of a quantum critical point, whose nature has not yet been identified. 
Here we present calculations for the two-dimensional Hubbard model in the doped Mott insulator regime, which indicate that the specific heat anomaly can arise from the finite temperature critical endpoint of a first-order transition between a pseudogap phase with dominant singlet correlations and a metal. As a function of doping at the temperature of the endpoint, the specific heat diverges. Upon increasing temperature, the peak becomes broader. The diverging correlation length is associated with uniform density fluctuations. No broken symmetries are needed.  
These anomalies also occur at half-filling as a function of interaction strength, and are relevant for organic superconductors and ultracold atoms.
\end{abstract}

\maketitle

{\it Introduction.--} 
The pseudogap phase in hole-doped cuprate superconductors indicates a partial loss of low energy excitations. The temperature and doping dependent boundary $T^*(\delta)$ of the pseudogap is seen in many physical properties, but whether it is a phase transition or a crossover is not always clear~\cite{keimerRev,normanADV}.

At the doping $\delta^*$ where the pseudogap ends there is a confluence of several phenomena such as robust superconductivity, linear dependence on temperature of resistivity~\cite{Taillefer:AnnuRev2018}, and divergent normal-state electronic specific heat $C$ divided by temperature $T$, $C/T$, versus doping~\cite{Michon:Cv2018}. It has thus been proposed that $\delta^*$ may represent a quantum critical point between competing phases, in analogy with heavy-fermion systems~\cite{RoschRMP:2007} and iron-based superconductors~\cite{Walmsley:2013}. However, the nature of the broken symmetry state giving rise to the pseudogap has not been clearly identified since in hole-doped cuprates a divergent correlation length associated to a broken symmetry state has not been found. The pseudogap may host many broken symmetry phases~\cite{keimerRev,Zhao:2017}. 

On the other hand, it has been proposed that the pseudogap can emerge upon doping the Mott insulator without invoking broken symmetry states: Mott localisation and short-range antiferromagnetic correlations form singlet bonds that open a pseudgap~\cite{Anderson:1987}, whose onset is marked by crossovers in thermodynamic quantities~\cite{Alloul:1989,keimerRev}. From a theoretical perspective, over the years the numerical solutions of the two-dimensional Hubbard model relevant for these systems support the idea that pseudogap originates from strong correlations~\cite{st,kyung,hauleDOPING,michelCFR,ssht,Wu:PRX2018}. 

How then to interpret thermodynamic anomalies such as divergent specific heat at $\delta^*$ where the pseudogap ends~\cite{Michon:Cv2018}? Is quantum criticality the only possible explanation for the specific heat anomaly? Here we answer these questions. We show that a specific heat anomaly is not necessarily a signature of a quantum critical point. It can occur because of the low-temperature critical endpoint of a first-order transition. This situation is not unique, as demonstrated by the divergent specific heat at the endpoint of the two-dimensional Ising model or at the liquid-gas endpoint in water. For cuprates, such an endpoint emerges from the normal-state solution of the two-dimensional Hubbard model in the doped Mott insulator regime~\cite{sht,sht2,ssht}. This mechanism solves the puzzle of a diverging correlation length at the doping level where the pseudogap ends, {\it without} the need of broken symmetry states.

{\it Model and method.--} 
The results reported in this Rapid Communication are based on the two-dimensional Hubbard model on the square lattice, $H=-\sum_{\langle ij\rangle \sigma}t_{ij} c_{i\sigma}^\dagger c_{j\sigma}  +U\sum_{i} n_{i\uparrow } n_{i\downarrow }  -\mu\sum_{i\sigma} n_{i\sigma},$ where  $t_{ij}=t$ is the nearest neighbor hopping, $U$ is the onsite Coulomb repulsion, $\mu$ is the chemical potential, $c^{\dagger}_{i\sigma}$ and $c_{i\sigma}$ operators create and annihilate an electron of spin $\sigma$ on site $i$, and $n_{i\sigma}=c^{\dagger}_{i\sigma}c_{i\sigma}$ is the number operator. We solve this model within the cellular extension~\cite{maier, kotliarRMP, tremblayR} of dynamical mean-field theory~\cite{rmp} (CDMFT). CDMFT extracts a cluster out of the lattice - here a $2\times 2$ plaquette - and replaces the missing lattice environment with a self-consistent bath of noninteracting electrons. We solve the resulting cluster in a bath problem using the continuous-time quantum Monte Carlo method~\cite{millisRMP} based on the hybridization expansion of the impurity action (CT-HYB).
In the nearest-neighbor square lattice model that we study, the Mott transition is hidden by long-range antiferromagnetic order~\cite{LorenzoAF, LorenzoAF2, Shafer2015, trilex2}. A model with frustrated antiferromagnetism would lead to the decrease of the antiferromagnetic transition temperature and would show the Mott transition. However, such a model leads to increased fermion sign problems, so we work with the simpler model. Thus our low-temperature results aim to give qualitative and not quantitative understanding of  experimental phenomena. In the following calculations, we do not allow symmetry breaking, so Mott physics is visible at all temperatures. Similarly, stripe order and other possible charge orders away from half-filling~\cite{Devereaux2018, Devereaux2018b} are not allowed in our calculation.

\begin{figure}
\centering{
\includegraphics[width=1.\linewidth]{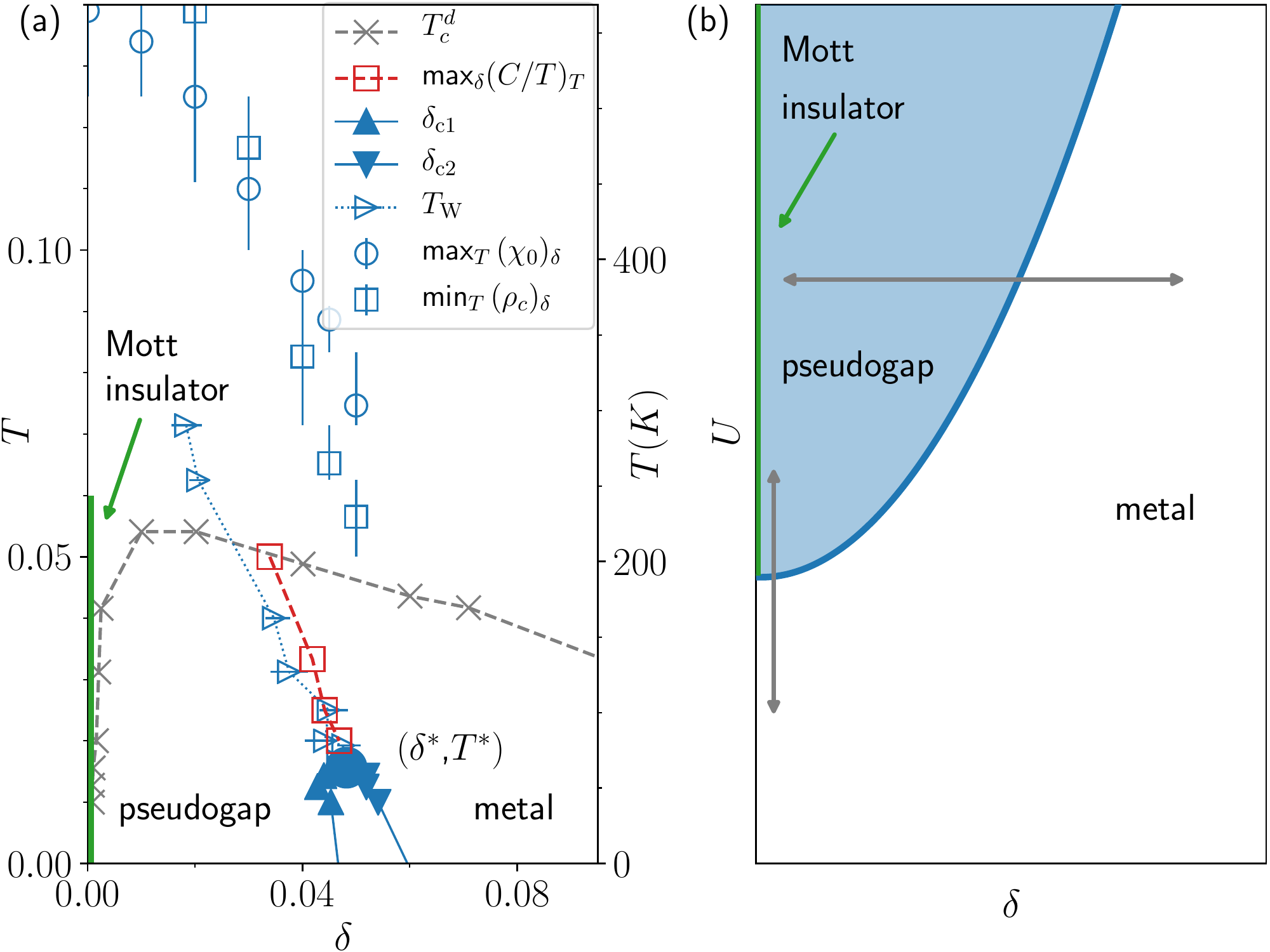}
}
\caption{Normal-state phase diagram of the two-dimensional Hubbard model within plaquette CDMFT. (a) Temperature versus doping phase diagram for $U=6.2t>U_{\rm MIT}$~\cite{sht2,ssht,sshtRHO,CaitlinOpalescence}. At zero doping there is a Mott insulator. At finite doping there is a first-order transition between a pseudogap phase and a metal. This first-order transition is bounded by the spinodal lines $\delta_{c1}$ and $\delta_{c2}$ and terminates at the critical endpoint $(\delta^*,T^*)$. From the endpoint emerges the Widom line, $T_W$, here defined as the locus of the maxima of isothermal charge compressibility $\kappa_T$ as a function of doping~\cite{ssht}. The red squares indicate the loci of the maximum in specific heat as a function of doping at constant temperature. This is one of our key findings. The open circles and squares denote extrema in, respectively, spin susceptibility (from Ref.~\cite{ssht}) and $c$-axis resistivity (from Ref.~\cite{sshtRHO}). Only below $\delta^*$, does the Widom line have a high-temperature precursor, where spin susceptibility drops vs $T$ and $c$-axis resistivity rises vs $T$~\cite{sshtRHO}. These indicators are often used to mark the pseudogap temperature $T^*(\delta)$. Crosses indicate the dynamical mean-field superconducting transition temperature $T_c^d$ (from Ref.~\cite{LorenzoSC}). On the right vertical axis we convert into physical units by using $t=350$meV. (b) Sketch of the interaction strength versus doping phase diagram at low temperature. The blue line indicates the first-order transition extending from $\delta=0$~\cite{sht2}. At $\delta=0$, for increasing $U$ it separates a metal from a Mott insulator (vertical green line). The shaded blue region is the pseudogap, which is an emergent phase that occurs in the doped Mott insulator.  
Horizontal (vertical) double arrow denotes the doping-driven (interaction-driven) Mott transition. 
}
\label{fig1}
\end{figure}
%

{\it Phase diagram of the two-dimensional Hubbard model.--} 
Hole-doped cuprates are doped Mott insulators. Hence we model these systems with the two-dimensional Hubbard model with nearest-neighbor hopping $t$. Figure~\ref{fig1}(a) shows the $T-\delta$ phase diagram of this model solved with plaquette cellular dynamical mean-field theory, for a value of the interaction strength $U=6.2t$, which is larger than the critical threshold $U_{\rm MIT}$ necessary to open a Mott insulator at half-filling. As a result of intense scrutiny~\cite{sht2,ssht,sshtRHO}, this phase diagram is known. There is a first-order transition at finite doping and finite temperature between a correlated metal at high doping and a strongly correlated pseudogap phase with predominant singlet correlations at low doping. These phases have the same symmetry and differ in their electronic densities at the first-order transition~\cite{sht, sht2}. 
The importance of singlets in the pseudogap phase can be inferred either from direct measurements of singlet correlations on the plaquette~\cite{sht, sht2, ssht, hauleDOPING, gullEPL, michelEPL, michelPRB} or from the fall in the uniform spin susceptibility as temperature decreases~\cite{ssht, Macridin:2006, hauleDOPING, Chen:2019, Reymbaut:2019}. 
The first-order transition moves progressively towards larger doping levels and low temperatures with increasing $U$. It ends in a critical endpoint at $(\delta^*, T^*)$. From the endpoint, a crossover, $T_W(\delta)$, with the features of the so-called Widom line~\cite{water1,supercritical}, emerges~\cite{ssht}.  
The maxima of thermodynamic response functions at constant temperature [such as isothermal charge compressibility (open triangles)] converge to that line asymptotically close to the endpoint. It has a high-temperature precursor, as indicated by maxima of spin susceptibility (open circles~\cite{ssht}) and minima of $c$-axis resistivity (open squares~\cite{sshtRHO}) versus $T$. This high-temperature precursor of $T_W(\delta)$ would then be associated with the pseudogap temperature $T^*(\delta)$ in cuprates, as observed for instance by NMR Knight shift~\cite{Alloul:1989}.  

Figure~\ref{fig1}(b) shows a sketch of the $U-\delta$ phase diagram at low temperature.  
First, the transition between pseudogap and correlated metal is connected, in the $U-\delta$ plane, to the metal to Mott insulator transition at zero doping. This implies~\cite{sht} that the pseudogap-metal transition originates from Mott physics and short-range correlations. Second, the first-order transition moves to progressively larger doping as $U$ increases. The latter point has important implications when comparing the phase diagram of Fig.~\ref{fig1}(a) to the experimental phase diagram of cuprates. $U=6.2t$ produces a gap at half-filling of order $0.45$eV~\cite{ssht}, whereas typical experimental values are of order $2$eV. Hence, to have a Mott insulating gap as found in experiments, one needs a value of $U$ about $9t-12t$. In that case the pseudogap would end around $\delta \approx 0.12$~\cite{hauleDOPING,sht2}. The sign problem prevents us from directly accessing regions with large values of $U$ and low $T$. As a consequence, one must keep in mind that in applying our model to cuprates, the doping at which the pseudopgap phase ends is smaller by about a factor 4 when compared to experiment.  Furthermore, bandstructure effects taken into account by next nearest-neighbor hopping are not considered here, and further contribute to push the doping at which the pseudogap ends towards larger doping. 
The experimental relevance of these findings can be found in Refs.~\cite{ssht,sshtRHO,Alloul2013}.

\begin{figure}
\centering{
\includegraphics[width=1.\linewidth]{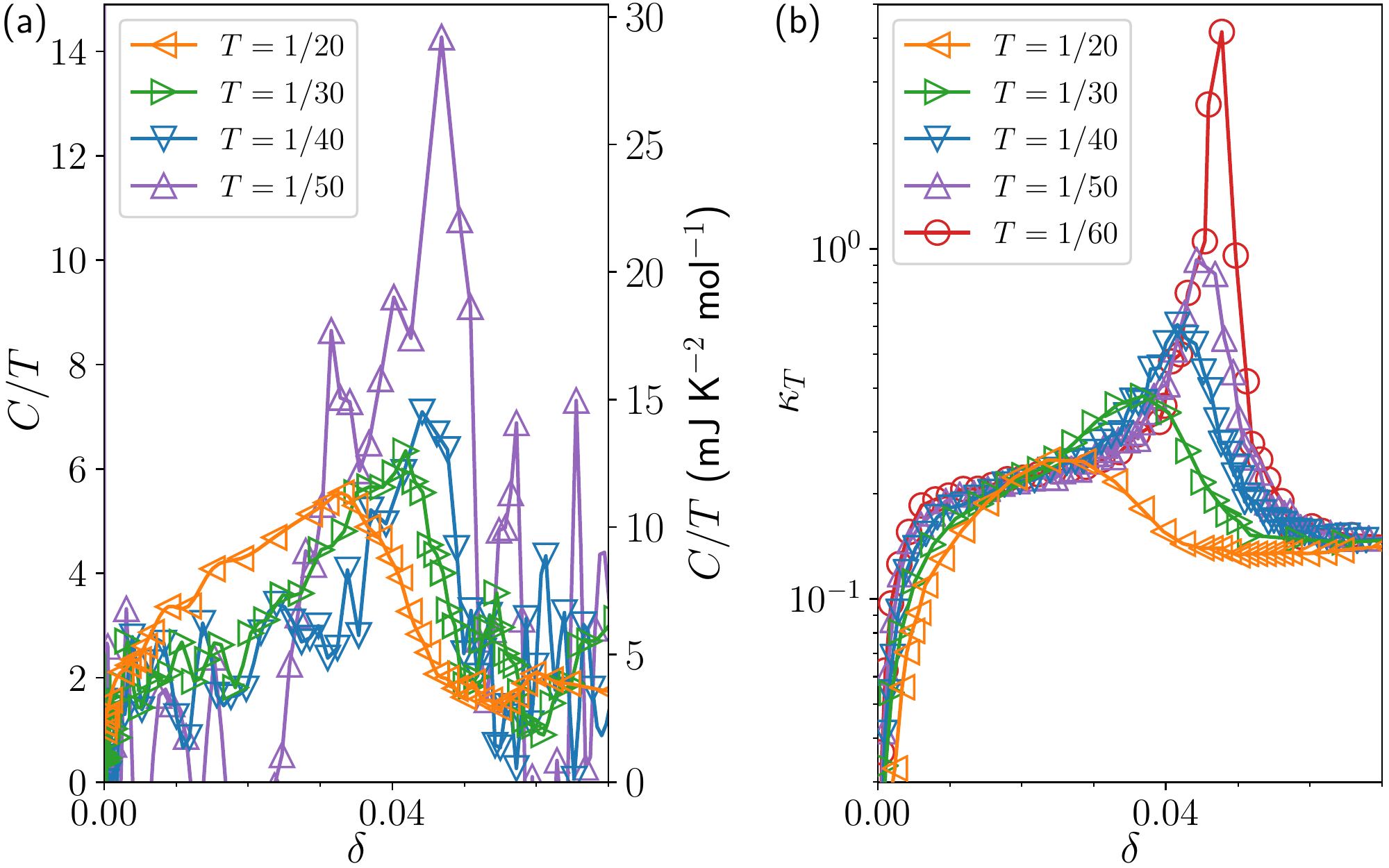}
}
\caption{(a) $C/T=(\partial A/ \partial T)/T$ versus hole doping $\delta$ for different temperatures above $T^*$ for $U=6.2t>U_{\rm MIT}$. We perform numerical derivative by finite differences between two temperatures. $C/T$ at $T=1/20, 1/30, 1/40, 1/50$ is evaluated by taking finite difference between $T=1/20$ and $T=1/30$, between $T=1/30$ and $T=1/40$, between $T=1/40$ and $T=1/50$, and between $T=1/50$ and $T=1/60$,  respectively. Maximum of $C/T$ at each temperature is shown in Fig.~\ref{fig1}(a) with red squares. On the right vertical axis we convert into physical units by using $t=350$meV. (b) Isothermal charge compressibility $\kappa_T=1/n^{2} (dn/d\mu)_T$ versus doping for different temperatures. Maximum of $\kappa_T$ at each temperature is shown in Fig.~\ref{fig1}(a) with open triangles.
}
\label{fig2}
\end{figure}
%

{\it Specific heat versus doping.--} 
We now turn to the behavior of the specific heat. First we calculate the thermodynamic potential $A=E_{kin}+E_{pot}-\mu n$~\footnote{The filling $n$ and the double occupancy $D$ needed for $E_{pot}=UD$ are calculated directly within the continuous-time quantum Monte Carlo impurity solver. As for the calculation of the kinetic energy $E_{kin}$, we have used the methodology introduced in Ref.~\onlinecite{LorenzoSC}. In short, the kinetic energy can be written as the sum of two contributions: the first one is related to the average expansion order term in the CT-HYB impurity solver, and the second one is coming from the cluster part and can be computed from the cluster density matrix.}, and then we perform a numerical derivative to extract the specific heat $C=(\partial A/\partial T)_\mu=T(\partial S/\partial T)_\mu$, which is a good approximation for $C=T(\partial S/\partial T)_\delta$ when the thermal expansion coefficient $(\partial V/\partial T)_\mu$ can be neglected, which is usually the case at low temperature because of the third law of thermodynamics.  

Figure~\ref{fig2}(a) shows $C/T$ as a function of doping $\delta$ for several temperatures. Deep in the Mott insulating state at zero doping and low temperature the $C/T$ is zero. At finite doping, $C/T$ exhibits a peak as a function of $\delta$. This is compatible with the experimental rise of $C/T$~\cite{Loram:1993, Michon:Cv2018} vs doping, followed by a drop at $\delta^*$~\cite{Michon:Cv2018}. 

Upon lowering $T$ towards the pseudogap endpoint $(T^*, \delta^*)$, the position of the maximum as a function of doping moves to higher doping, the peak sharpens and its magnitude increases. The position of the specific heat maxima are shown in Fig.~\ref{fig1}(a) with red squares.  
At the endpoint $C/T$ diverges. Noise associated to numerical derivative prevents us from approaching the endpoint to capture the divergence. 
Nevertheless, thermodynamic anomalies, such as those in $C/T$, occur in {\it any} thermodynamic variable~\cite{water1}. Figure~\ref{fig2}(b) shows for instance the charge compressibility $\kappa_T=1/n^2 dn/d\mu$ as a function of doping $\delta$ for several temperatures. The divergence of $\kappa_T$ at the endpoint is clearly visible.  
Charge compressibility is a $q=0$ quantity. Various pseudogap signatures at $q=0$ are observed e.g. with scanning tunneling microscopy~\cite{Fujita:2014} and neutrons~\cite{Fauque:2006}. Fig.~\ref{fig1}(a) shows that the locus of specific heat maxima follows the position of charge compressibility maxima ($T_{W}(\delta)$). 

Two comments are in order. First, we are working with a minimal theoretical model, so we look for qualitative and not quantitative agreement with experiments. Larger values of $U$ push the doping at which the specific heat diverges, or more generally where the pseudopgap phase ends, $\delta^*$, at higher doping, as demonstrated in Refs.~\onlinecite{sht2, LorenzoSC}. Larger values of $U$ also push the critical endpoint to much lower temperatures.
Second, the peak in $C/T$ is not due to the renormalized van Hove singularity~\cite{Wu:PRX2018,Reymbaut:2019}, which occurs at larger doping and is essentially temperature independent~\cite{sht2}, in sharp contrast with the marked temperature dependence of $T^*(\delta)$. 
Calculations with different values of frustration confirm the distinction between pseudogap endpoint and van Hove singularity~\cite{Wu:PRX2018}. Experimentally, the peak in the electronic specific heat is not due to the van Hove singularity~\cite{Michon:Cv2018,Horio:PRL2018}.

{\it Discussion.--} 
Our results provide a coherent microscopic theoretical model to understand what may occur in hole-doped cuprates.  
First, the peak in $C/T$ as a function of doping is a signature of a critical endpoint, which can be confused with a quantum critical point since it occurs at low temperature for large $U$. However, the crossover arising from the endpoint has three distinct signatures that can be experimentally tested: (a) The peak in $C/T$ becomes broader with raising temperature. This is compatible with experiments~\cite{Michon:Cv2018}. (b) At temperatures below the endpoint, there is a first-order transition where the peak in $C/T$ disappears. The Clausius-Clapeyron relation implies that entropy of the pseudogap is smaller than the entropy of the metal at larger doping. (c) At the critical endpoint, critical scaling is expected. This may provide an alternate explanation for the $-\ln T$ behavior when the temperature of the critical endpoint is small enough~\cite{Bartosch:2010} (in mean-field we have the exponent $\alpha=0$). 

Second, the interpretation of a diverging $C/T$ as a signature of a quantum critical point is complicated by the fact that no symmetry broken state ends at $\delta^*$, where the pseudogap ends. As pointed out in Ref.~\cite{Michon:Cv2018}, no diverging antiferromagnetic correlation length and no diverging spin density wave correlation length occur near $\delta^*$. In our calculations, the diverging $C/T$ marks the endpoint separating phases with the same symmetries, as in the liquid-gas transition. Pseudogap and metal have different electronic densities at the first-order transition. At the endpoint the diverging correlation length is associated with density fluctuations~\cite{CaitlinOpalescence}. Correspondingly, enhancement of specific heat is associated with large energy fluctuations. 

Third, the interpretation of a quantum critical point is at odds with the experimental finding of an abrupt end of the pseudogap temperature at $\delta^*$ with a sizeable finite $T^*$~\cite{Collignon_Badoux_Taillefer:2017,Olivier:PRB2018}. In our theoretical model, this abrupt end of the pseudogap temperature occurs because both the pseudogap and its associated crossovers end at a finite-temperature first-order transition. Our Ref.~\cite{sshtRHO} already suggested that $T^*$ should not be extrapolated to $T=0$. Ref.~\cite{Doiron-Leyraud_Cyr-Choiniere_Taillefer_2017} suggests that the abrupt fall comes from the constraint that $\delta^*$ is less or equal to the doping where there is a van Hove singularity.  
 
Fourth, our calculations suggest a possible explanation for the proliferation of long-range or quasi long-range ordered phases detected near $\delta^*$~\cite{keimerRev}. Large charge and energy fluctuations along the Widom line may develop into charge density ordered phases~\cite{keimerRev,Fujita:2014}. Superconductivity in our calculations straddles the critical  endpoint~\cite{LorenzoSC}, similarly to experiments.

\begin{figure}
\centering{
\includegraphics[width=1.\linewidth]{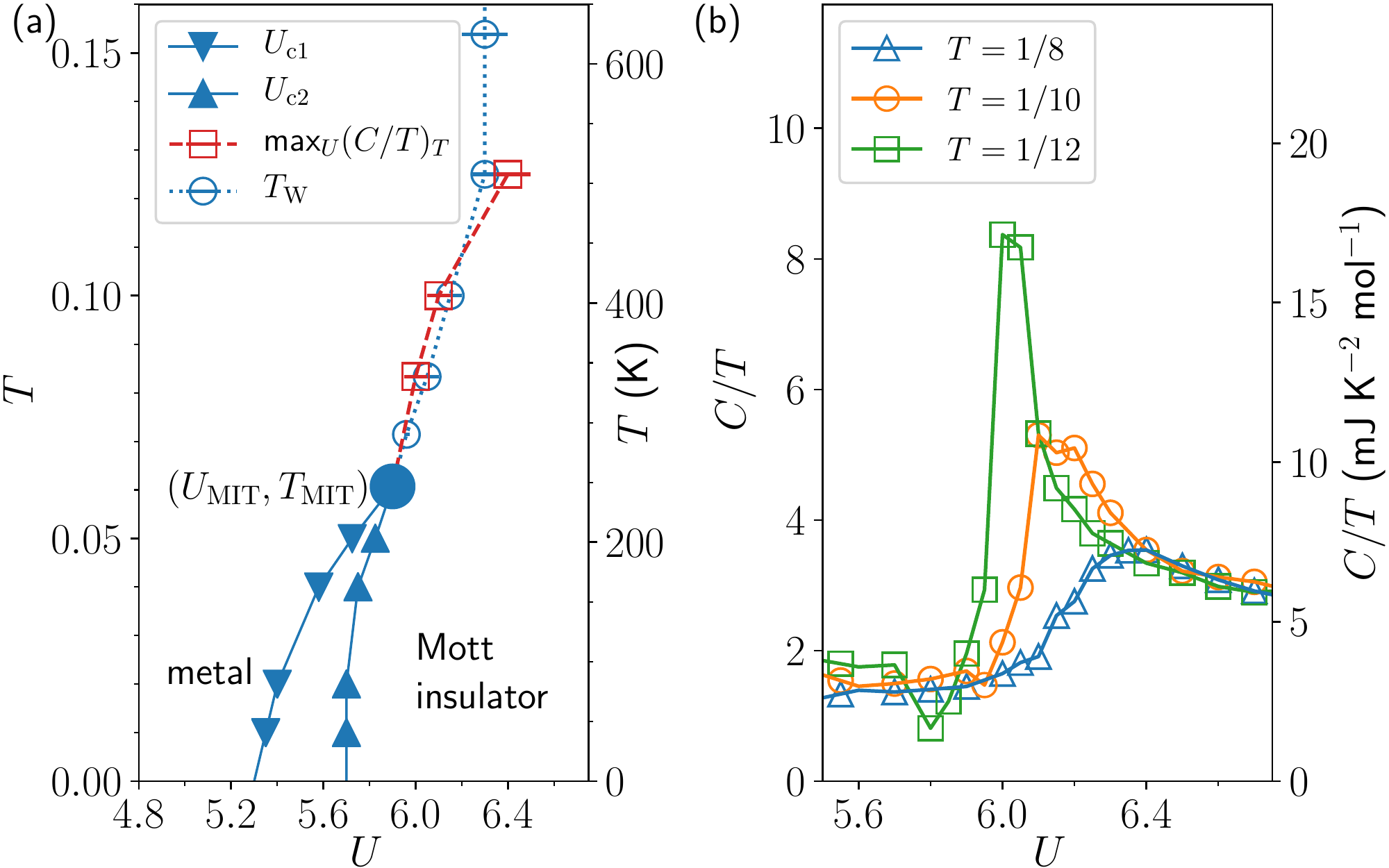}
}
\caption{(a) Temperature versus interaction strength normal-state phase diagram of the two-dimensional Hubbard model at half filling ($\delta=0$) within plaquette CDMFT~\cite{CaitlinSb}. The first-order transition between a metal and a Mott insulator is bounded by the spinodal lines $U_{c1}$ and $U_{c2}$, and ends at the Mott endpoint $(U_{\rm MIT}, T_{\rm MIT})$. The Widom line $T_W$ is estimated from the inflection point in the double occupancy $D(U)_T$ (from Ref.~\cite{CaitlinSb}). (b) $C/T=(\partial A/\partial T)/T$ versus $U$ for different temperatures. We perform numerical derivatives by finite differences between two temperatures. $C/T$ at $T=1/8, 1/10, 1/12$ is evaluated by taking finite difference between $T=1/8$ and $T=1/10$, between $T=1/10$ and $T=1/12$, and between $T=1/12$ and $T=1/14$, respectively. Maximum of $C/T$ as a function of $U$ at each temperature is shown in (a) with red squares. On the right vertical axis we convert into physical units by using $t=350$meV. 
}
\label{fig3}
\end{figure}
%

{\it Specific heat versus interaction strength.--} 
Enhancement of $C/T$ at an endpoint has implications beyond the physics of cuprates. Specific heat maxima occur along the pseudogap to metal transition in the $U-\delta$ diagram of Fig.~\ref{fig1}(b). This pseudogap to metal transition is connected to the metal to Mott insulator transition at zero doping. The latter is relevant for the physics of organic superconductors~\cite{Lefebvre:2000,Bartosch:2010} and for some transition-metal oxides~\cite{ift}, and can be simulated with ultracold atoms in optical lattices~\cite{GrossScience2017,bdzRMP,jzHM,lewenstein,Esslinger:2010}. Figure~\ref{fig3}(a) shows the $T-U$ phase diagram of the two-dimensional Hubbard model 
with plaquette CDMFT at half-filling. This phase diagram is known~\cite{phk,CaitlinSb}: there is a first-order transition terminating at the critical endpoint $(U_{\rm MIT}, T_{\rm MIT})$ out of which emerges a Widom line (here defined as the loci of inflection in the double occupancy versus $U$~\cite{CaitlinSb}). Contrary to the previous case, at half-filling the first-order transition separates a metal from a Mott insulator. 

Figure~\ref{fig3}(b) shows $C/T$ as a function of $U$ for several temperatures. Again, $C/T$ shows a peak that narrows and whose intensity increases when approaching the temperature of the endpoint $T_{\rm MIT}$ from above. 
Below the Mott endpoint, $C/T$ is discontinuous and the Clausius-Clapeyron relation implies that the entropy of the insulator is smaller than that of the metal. 
Our results are consistent with the rapid decrease of $C/T(U)$ in layered organic conductors for $U>U_{\rm MIT}$~\cite{Nakazawa:2000}. They are also compatible with the increase of $C/T(U)$ in a 2D ${}^{3}$He fluid monolayer for $U<U_{\rm MIT}$~\cite{Casey:2003}.

{\it Conclusions.--} 
In summary, using a two-dimensional Hubbard model, we unveiled a mechanism in which one can rationalise the thermodynamic anomalies in hole-doped cuprate superconductors and organic superconductors. They may not reflect the presence of a quantum critical point. They are instead caused by a critical endpoint at very low temperature that arises from Mott physics plus short-range singlet correlations.

\section*{Acknowledgments}
We thank L. Taillefer and J. Saunders for discussions. This work has been supported by the Natural Sciences and Engineering Research Council of Canada (NSERC) under grants RGPIN-2014-04584, the Canada First Research Excellence Fund and by the Research Chair in the Theory of Quantum Materials. PS work was supported by the U.S. Department of Energy, Office of Science, Basic Energy Sciences as a part of the Computational Materials Science Program. Simulations were performed on computers provided by the Canada Foundation for Innovation, the Minist\`ere de l'\'Education des Loisirs et du Sport (Qu\'ebec), Calcul Qu\'ebec, and Compute Canada.


%

\end{document}